\documentclass[runningheads]{llncs}
\usepackage{amsmath,amssymb,amsfonts}
\usepackage{marvosym}
\usepackage{booktabs} 
\usepackage[T1]{fontenc}
\usepackage{multirow}
\usepackage{cite, hyperref, tikz}
\usepackage{hyperref}
\usepackage{url}
\hypersetup{
	colorlinks=true,
	linkcolor=cyan,
	filecolor=blue,      
	urlcolor=black,
	citecolor=green,
}
\usepackage{colortbl}
\hypersetup{
            colorlinks=true,
            linkcolor=blue,
            anchorcolor=blue,
            citecolor=blue
            }

\usepackage{graphicx}
\begin{document}
\title{Cross-Organ and Cross-Scanner Adenocarcinoma Segmentation using Rein to Fine-tune Vision Foundation Models}
%
%
\author{Pengzhou Cai\inst{1,2} \and
Xueyuan Zhang\inst{2} \and
Libin Lan\textsuperscript{(\Letter)}\inst{1} \and
Ze Zhao\textsuperscript{(\Letter)}\inst{3}}
\authorrunning{P. Cai et al.}
%
\institute{Chongqing University of Technology, Chongqing 400054, China \\
\email{lanlbn@cqut.edu.cn}
\and
Chongqing Zhijian Life Technology Co. LTD, Chongqing 400039, China \\
\email{\{caipzh, zhangxy\}@zhijianlife.cn}
\and
Institute of Computing Technology, Chinese Academy of Sciences, Beijing 100190, China \\
\email{zhaoze@ict.ac.cn}}

\maketitle              
\begin{abstract}
In recent years, significant progress has been made in tumor segmentation within the field of digital pathology. However, variations in organs, tissue preparation methods, and image acquisition processes can lead to domain discrepancies among digital pathology images. To address this problem, in this paper, we use Rein, a fine-tuning method, to parametrically and efficiently fine-tune various vision foundation models (VFMs) for MICCAI 2024 Cross-Organ and Cross-Scanner Adenocarcinoma Segmentation (COSAS2024). The core of Rein consists of a set of learnable tokens, which are directly linked to instances, improving functionality at the instance level in each layer. In the data environment of the COSAS2024 Challenge, extensive experiments demonstrate that Rein fine-tuned the VFMs to achieve satisfactory results. Specifically, we used Rein to fine-tune ConvNeXt and DINOv2. Our team used the former to achieve scores of 0.7719 and 0.7557 on the preliminary test phase and final test phase in task1, respectively, while the latter achieved scores of 0.8848 and 0.8192 on the preliminary test phase and final test phase in task2. Code is available at \href{https://github.com/ZhiJianLife/cosas2024_Zhijian-Life}{GitHub}.

\end{abstract}
\section{Introduction}
Adenocarcinomas are prevalent in tissues such as the breast, stomach, colon, and lungs, making glandular segmentation, particularly of adenocarcinoma regions, a primary focus. In recent years, key challenges in adenocarcinoma segmentation, such as (GlaS \cite{glas}, DigestPath \cite{digestpath}), have driven advances in digital pathology, significantly improving tumor diagnosis and localization. However, the inherent variability in digital pathology images and tissue types has posed substantial challenges for existing algorithms. Differences in organs, tissue preparation methods, and image acquisition processes have led to what is known as domain shifting. To address this issue, MIDOG 21/22 \cite{mitosis21, mitosis22} have developed. In order to further improve the generalization ability of segmentation model in the field of pathological images, we utilize Rein \cite{rein} to fine-tune the VFMs (e.g., Segment Anything Model (SAM) \cite{sam}, ConvNeXt \cite{convnext}, DINOv2 \cite{dinov2}) for MICCAI 2024 Cross-Organ and Cross-Scanner Adenocarcinoma Segmentation \footnote{https://cosas.grand-challenge.org/ \label{cosas}}. Our contributions are as follows:

$\bullet$ We utilize Rein to fine-tune the SAM, ConvNeXt, DINOv2 for MICCAI 2024 COSAS2024 Challenge, in which Rein is a fine-tuning method. The core of Rein consists of a set of learnable tokens, which are directly linked to instances, improving functionality at the instance level in each layer.

$\bullet$For task1, our team achieved scores of 0.7719 and 0.7557 on the preliminary test phase and final test phase in task1, respectively. For task2, our team achieved scores of 0.8848 and 0.8192 on the preliminary test phase and final test phase in task2, respectively.
\begin{figure*}
\centering
\includegraphics[width=1.0\linewidth, keepaspectratio]{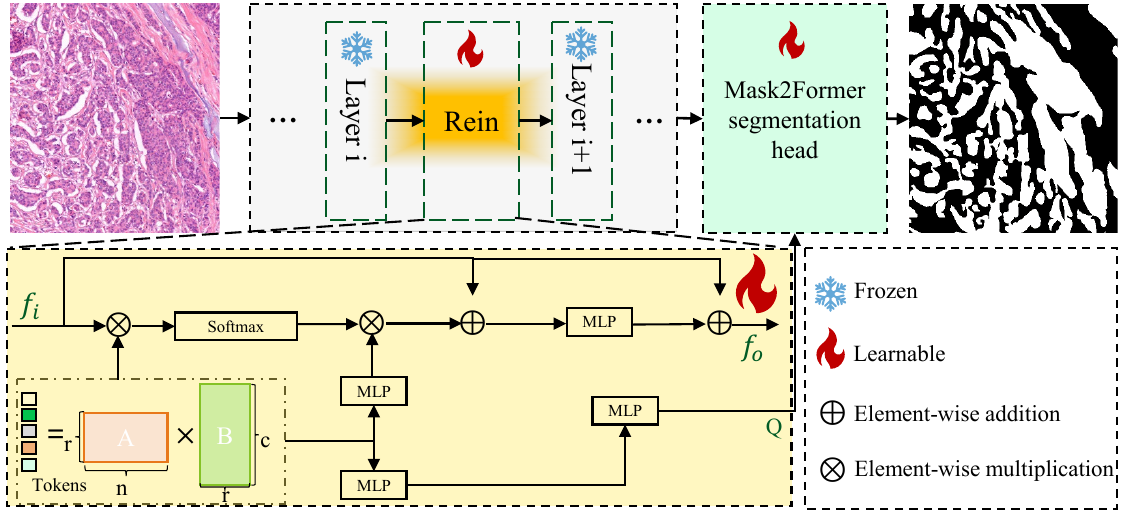}
\caption{The network architecture of the method. We freeze the weights for backbone and fine-tune the weights for the rein module and Mask2Former. $f_i$ and $f_o$ represent the output of the previous layer and the input of the next layer respectively. $Q$ is the object of the token map, and Q is then linked to the instance to achieve enhanced performance.}
\label{fig1:net}
\end{figure*}
\section{Methods}
In the field of Natural Language Processing (NLP), parameter-efficient fine-tuning (PEFT) has proven highly effective by keeping most VFMs parameters frozen and fine-tuning only a small subset. The \cite{rein} proposed embedding a mechanism, named 'Rein', between the layers of the backbone. Rein actively refines and passes feature maps from each layer to the next, allowing for more effective utilization of VFMs' powerful capabilities. Inspired by \cite{rein}, we utilize Rein to fine-tune the SAM, ConvNeXt, DINOv2 for MICCAI 2024 COSAS2024 Challenge. Fig. \ref{fig1:net} shows the network architecture of the method. Specifically, the network contains encoder and decoder. For encoder, freezing backbone and fine-tuning Rein is a key step in incorporating medical knowledge into the vision foundation model. For decoder, we adopt Mask2Former \cite{mask2former} as our segmentation head because it integrates various VFM as the backbone. The core of Rein consists of a set of learnable tokens, which are directly linked to instances, improving functionality at the instance level in each layer. To significantly reduce the number of parameters, similar to LoRA \cite{lora}, it implemented low-rank token sequence. Moreover, to address the redundancy of parameters in a layer-specific multilayer perceptron (MLP) weight, a shared MLP weight is used between the layers.
\begin{table}
\centering
\caption{Details of the dataset of the COSAS2024 Challenge. 
 (I/C) represents the number of images and the number of categories (organs or scanners).  }
\label{tab1:data}
\resizebox{1.0\linewidth}{!}
{
\begin{tabular}{l|ccccc}
\toprule
COSAS2024 &Image Size & Total(I/C) & Train(I/C) & Preliminary test (I/C) & Final test(I/C) 
\\
\midrule
Task1 &1500$\times$1500 & 290/6 & 180/3 & 20/4 & 90/6
\\
Task2  &1500$\times$1500 & 290/6 & 180/3 & 20/4 & 90/6 
\\

\bottomrule
\end{tabular}
}
\end{table}
\section{Experiment}
\subsection{Dataset}
The dataset of COSAS2024 \footnote{https://cosas.grand-challenge.org/datasets/ \label{cosasdata}} is the first and largest domain generalization dataset for the digital pathology segmentation task. There are two factors causing the domain shift: different organs in task1 and different scanners in task2.

For task1, the dataset consisted of 290 pathological images of six different adenocarcinomas extracted from the WSI digitized by the TEKSQRAY SQS-600P scanner, with an average size of 1500 x 1500 pixels. The training set consisted of 180 images from 3 organs (gastric adenocarcinoma, colorectal adenocarcinoma and pancreatic ductal adenocarcinoma), the preliminary test set consisted of 20 images from 4 organs, and the final test set consisted of 90 images from 6 organs.

For task2, the data included 290 images of invasive breast cancer obtained from six different WSI scanners, each approximately 1500 x 1500 pixels in size. The training set consisted of 180 images from 3 scanners, the preliminary test set consisted of 20 images from 4 scanners, and the final test set consisted of 90 images from 6 scanners. Please refer to Table \ref{tab1:data} for detailed data.

\subsection{Implementation details}
The method is implemented based on MMsegmentation \cite{mmsegmentation} codebase.  All our experiments are conducted on a single NVIDIA GeForce RTX 4090 with 24GB. We chose three VFMs as fine-tuning objects, including SAM-h, ConvNeXt, DINOv2. For the training pharse, we employed the AdamW optimizer to optimize the model during the back propagation. We configured the learning rate to 1e-5 for the backbone and 1e-4 for both the decode head and the Rein. In addition, we use a setup of 60,000 iterations with a batch size of 4, cropping images to a resolution of 512 × 512.

\section{Results}
\begin{figure}
\centering
\includegraphics[width=1.0\linewidth, keepaspectratio]{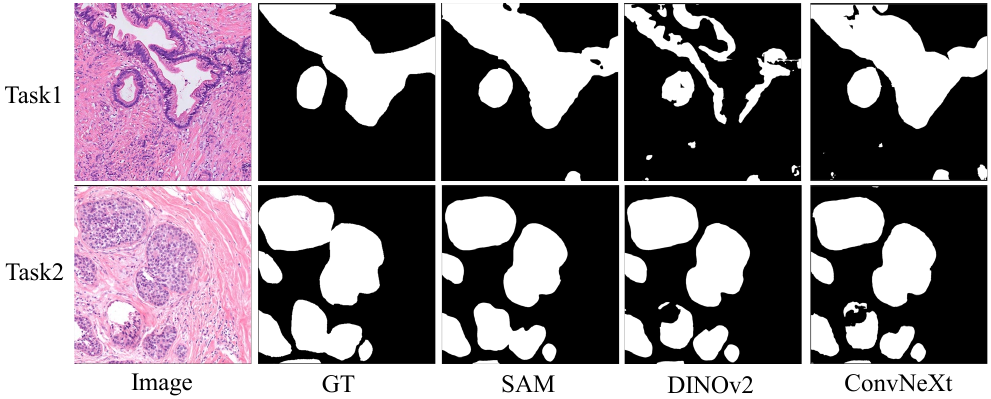}
\caption{The segmentation results of different methods for the dataset of both tasks.  }
\label{fig4:synapse}
\end{figure}

First of all, since the test set is not publicly available at present, in the training stage, we randomly divide the training set in a ratio of 8:2 for training and testing to evaluate the methods. To evaluate the segmentation performance of the different methods, we utilize two common evaluation metrics: average Dice-Similarity Coefficient (DSC), the mean Intersection over Union (mIoU). Table \ref{tab:pre} shows the segmentation results of different methods for task1 and task2. When combined with Table \ref{tab:pre} and Figure \ref{fig4:synapse} , it is clear that fine-tuning ConvNeXt and DINOv2 using rein for task1 and task2, respectively, achieves the best segmentation results and demonstrates strong generalization capabilities.

By detailing the features of each layer of backbone and connecting it to instances,  rein can greatly narrow the gap between different organs and scanner domains. Otherwise, due to the differences between the pre-training weights of different models, we found that ConvNext's pre-trained model was more suitable for cross-organ adenocarcinoma segmentation, while DINOv2's pre-trained model was more suitable for cross-scanner adenocarcinoma segmentation.

\begin{table}
\centering
\caption{The quantitative results of the different methods for the dataset of task1 and task2.  The symbol $\uparrow$ indicates the larger the better. The best result is in \textbf{Blod}. }
\begin{tabular}{c|cc|c|cc}
\hline
Task1  & DSC $\uparrow$ & mIoU $\uparrow$ & Task2 & DSC $\uparrow$ & mIoU $\uparrow$ \\ \hline
ConvNeXt & \textbf{0.8568}    &  0.7433  & ConvNeXt   &  0.8959  & 0.7497 \\ \hline
SAM & 0.8489   & \textbf{0.7502}  &  SAM &  0.8914  &  0.7665  \\\hline
DINOv2 & 0.8477   & 0.7426  &  DINOv2 &  \textbf{0.9054}  &  \textbf{0.7721}  \\\hline
\end{tabular}
\label{tab:pre}
\end{table}

In addition, we present the evaluation results on the preliminary test set and the final test set in Table \ref{tab:chubutest} and in Table \ref{tab:final}. The final score is expressed as: scores = 0.5$ \times $ DSC + 0.5 $\times$ JSC, where JSC shows Jaccard Similarity Coefficient. For task1, our team achieved a score of 0.7719 and 0.7557 on
the preliminary test phase and final test phase using Rein to fine-tune
ConvNeXt, respectively. For task2, our team achieved a score of 0.8848
and 0.8192 on the preliminary test phase and final test phase using Rein
to fine-tune DINOv2, respectively.
\begin{table}
\centering
\caption{The quantitative results of the different teams on the preliminary test set for the task1 and task2. We have only listed the results of the top ten teams and the pink areas in the table represent our team's results.}
\begin{tabular}{c|c|c|c}
\hline
Teams  & Task1 score $\uparrow$ &Teams & Task2 score$\uparrow$  \\ \hline
agalaran & 0.7865    &deepmicroscopy &  0.8858   \\ \hline
deepmicroscopy & 0.7776   &Zhijian Life\cellcolor{pink!40}& 0.8848\cellcolor{pink!40}  \\\hline
Zhijian Life\cellcolor{pink!40} & 0.7719\cellcolor{pink!40}   &ICT\_team& 0.8833   \\\hline
ICT\_team & 0.7714  & SMF&0.8775  \\\hline
Sanmed\_AI & 0.7633   &Biototem& 0.8728   \\\hline
Biototem & 0.7625   &DeepLearnAI& 0.8646   \\\hline
Amaranth & 0.7534  &Amaranth & 0.8643   \\\hline
baseline & 0.7351  &agalaran & 0.8584   \\\hline
DeepLearnAI & 0.7198   &Sanmed\_AI& 0.8457   \\\hline
excute(me) & 0.7049   &Team-Tiger&  0.8408  \\\hline
\end{tabular}
\label{tab:chubutest}
\end{table}
\begin{table}
\centering
\caption{The quantitative results of the different teams on the final test set for the task1 and task2. The pink areas in the table represent our team's results.}
\begin{tabular}{c|c|c|c}
\hline
Teams  & Task1 score $\uparrow$ &Teams & Task2 score$\uparrow$  \\ \hline
deepmicroscopy & 0.8020    &deepmicroscopy &  0.8527   \\ \hline
 ICT\_team& 0.7976   &Biototem & 0.8354   \\\hline
Amaranth & 0.7774   &Zhijian Life\cellcolor{pink!40} & 0.8192\cellcolor{pink!40} \\\hline
Sanmed\_AI & 0.7753  & agalaran&0.8175  \\\hline
Biototem & 0.7643   &Amaranth& 0.8128   \\\hline
agaldran & 0.7607   &Team-Tiger& 0.8093   \\\hline
Team-Tiger & 0.7583  &ICT\_team & 0.7944   \\\hline
Zhijian Life\cellcolor{pink!40} & 0.7557\cellcolor{pink!40}  &SMF & 0.7924   \\\hline
DeepLearnAI & 0.7469   &DeepLearnAI& 0.7597   \\\hline
Long Xin & 0.6446   &Sanmed\_AI&  0.7420  \\\hline
\end{tabular}
\label{tab:final}
\end{table}
\section{Conclusion}
In this paper, we use Rein to parameterically and efficiently fine-tune ConvNeXt and DINOv2 for MICCAI 2024 COSAS 2024. Extensive experiments demonstrate that Rein
fine-tuned the vision foundation model to achieve satisfactory results. We believe that Rein fine-tuned VFMs has great potential to generalize in the field of adenocarcinoma, whether adenocarcinoma comes from an organ or a scanner.

%
%
%

\begin{thebibliography}{10}
\bibitem{glas}
Korsuk Sirinukunwattana, Josien~PW Pluim, Hao Chen, Xiaojuan Qi, Pheng-Ann Heng, Yun~Bo Guo, Li~Yang Wang, Bogdan~J Matuszewski, Elia Bruni, Urko Sanchez, et~al.
\newblock Gland segmentation in colon histology images: The glas challenge contest.
\newblock {\em Medical image analysis}, 35:489--502, 2017.

\bibitem{digestpath}
Qian Da, Xiaodi Huang, Zhongyu Li, Yanfei Zuo, Chenbin Zhang, Jingxin Liu, Wen Chen, Jiahui Li, Dou Xu, Zhiqiang Hu, et~al.
\newblock Digestpath: A benchmark dataset with challenge review for the pathological detection and segmentation of digestive-system.
\newblock {\em Medical Image Analysis}, 80:102485, 2022.

\bibitem{mitosis21}
Marc Aubreville, Nikolas Stathonikos, Christof~A Bertram, Robert Klopfleisch, Natalie Ter~Hoeve, Francesco Ciompi, Frauke Wilm, Christian Marzahl, Taryn~A Donovan, Andreas Maier, et~al.
\newblock Mitosis domain generalization in histopathology images—the midog challenge.
\newblock {\em Medical Image Analysis}, 84:102699, 2023.

\bibitem{mitosis22}
Marc Aubreville, Nikolas Stathonikos, Taryn~A Donovan, Robert Klopfleisch, Jonas Ammeling, Jonathan Ganz, Frauke Wilm, Mitko Veta, Samir Jabari, Markus Eckstein, et~al.
\newblock Domain generalization across tumor types, laboratories, and species—insights from the 2022 edition of the mitosis domain generalization challenge.
\newblock {\em Medical Image Analysis}, 94:103155, 2024.

\bibitem{rein}
Zhixiang Wei, Lin Chen, Yi~Jin, Xiaoxiao Ma, Tianle Liu, Pengyang Ling, Ben Wang, Huaian Chen, and Jinjin Zheng.
\newblock Stronger fewer \& superior: Harnessing vision foundation models for domain generalized semantic segmentation.
\newblock In {\em Proceedings of the IEEE/CVF Conference on Computer Vision and Pattern Recognition}, pages 28619--28630, 2024.

\bibitem{sam}
Alexander Kirillov, Eric Mintun, Nikhila Ravi, Hanzi Mao, Chloe Rolland, Laura Gustafson, Tete Xiao, Spencer Whitehead, Alexander~C Berg, Wan-Yen Lo, et~al.
\newblock Segment anything.
\newblock In {\em Proceedings of the IEEE/CVF International Conference on Computer Vision}, pages 4015--4026, 2023.

\bibitem{convnext}
Zhuang Liu, Hanzi Mao, Chao-Yuan Wu, Christoph Feichtenhofer, Trevor Darrell, and Saining Xie.
\newblock A convnet for the 2020s.
\newblock In {\em Proceedings of the IEEE/CVF conference on computer vision and pattern recognition}, pages 11976--11986, 2022.

\bibitem{dinov2}
Maxime Oquab, Timoth{\'e}e Darcet, Th{\'e}o Moutakanni, Huy Vo, Marc Szafraniec, Vasil Khalidov, Pierre Fernandez, Daniel Haziza, Francisco Massa, Alaaeldin El-Nouby, et~al.
\newblock Dinov2: Learning robust visual features without supervision.
\newblock {\em arXiv preprint arXiv:2304.07193}, 2023.

\bibitem{mask2former}
Bowen Cheng, Ishan Misra, Alexander~G Schwing, Alexander Kirillov, and Rohit Girdhar.
\newblock Masked-attention mask transformer for universal image segmentation.
\newblock In {\em Proceedings of the IEEE/CVF conference on computer vision and pattern recognition}, pages 1290--1299, 2022.

\bibitem{lora}
Edward~J Hu, Yelong Shen, Phillip Wallis, Zeyuan Allen-Zhu, Yuanzhi Li, Shean Wang, Lu~Wang, and Weizhu Chen.
\newblock Lora: Low-rank adaptation of large language models.
\newblock {\em arXiv preprint arXiv:2106.09685}, 2021.

\bibitem{mmsegmentation}
MMSegmentation Contributors.
\newblock Mmsegmentation: Openmmlab semantic segmentation toolbox and benchmark.
\newblock \url{https://github.com/open-mmlab/mmsegmentation}, 2020.
\end{thebibliography}
%
\bibliographystyle{unsrt}

\end{document}